# Toward Substantive Intersectional Algorithmic Fairness: Desiderata for a Feminist Approach

**Mirsch**, Marie[1,*]; **Wegner**, Laila[2]; **Strube**, Jonas[1]; **Leicht-Scholten**, Carmen[1]

[1]Gender and Diversity in Engineering, RWTH Aachen University, Germany

[2]Philosophy & Ethics, Eindhoven University of Technology, Netherlands

*Corresponding author: marie.mirsch@gdi.rwth-aachen.de

**Abstract.** People's experiences of discrimination are often shaped by multiple intersecting factors, yet algorithmic fairness research rarely reflects this complexity. While intersectionality offers tools for understanding *how* forms of oppression interact, current approaches to intersectional algorithmic fairness tend to focus on narrowly defined demographic subgroups. These methods contribute important insights but risk oversimplifying social reality and neglecting structural inequalities. In this paper, we outline how a substantive approach to intersectional algorithmic fairness can reorient this research and practice. In particular, we propose *Substantive Intersectional Algorithmic Fairness*, extending Green's (2022) notion of substantive algorithmic fairness with insights from intersectional feminist theory. Aiming to provide as actionable guidance as possible, our approach is articulated as ten desiderata to guide the design, assessment, and deployment of algorithmic systems that address systemic inequities while mitigating harms to intersectionally marginalized communities. Rather than prescribing fixed operationalizations, these desiderata invite AI practitioners and experts to reflect on assumptions of neutrality, the use of protected attributes, the inclusion of multiply marginalized groups, and the transformative potential of algorithmic systems. By bridging computational and social science perspectives, the approach emphasizes that fairness cannot be separated from social context, and that in some cases, principled non-deployment may be necessary.

## 1 Introduction

When U.S. legal scholar Kimberlé Crenshaw coined[1] the term *intersectionality* in 1989 [2, 3], African-American women had long been advocating for recognition of their rights. Despite the progress of the U.S. civil rights, feminist, and labor movements of the 1960s and 1970s, these movements often failed to account for the compounded and structurally embedded forms of oppression faced by individuals who were *simultaneously* workers, Black[2], and female [4]. While these movements addressed Black women *implicitly*, they were predominantly driven by a single-axis lens on discrimination: being a worker, being Black, or being female. This focus on a single facet of being obscures the complex and multifaceted ways in which overlapping systems of oppression, such as racism, patriarchy, and classism, jointly shape the lived experiences of those multiply marginalized. Crenshaw [2] particularly critiques a recurring legal pattern she terms the *equality-difference*

[1] While we will frequently reference Crenshaw's work in this article since her structured critique of the legal justice system reveals numerous parallels to algorithmic systems, we want to emphasize two points: First, intersectionality's origins date back long before her article and have roots in activist movements, as we will highlight in Section 2. Second, discourses on intersectionality have evolved across multiple domains and contexts, producing a diverse and complex body of perspectives and theories. Despite this richness, intersectionality is often reductively associated solely with Crenshaw's name, an oversimplification that risks obscuring the collective activist efforts and erasing earlier epistemic contributions by Black women. As [1] aptly describes, „Crenshaw's name is appropriated and circulated as a decontextualized brand for intersectionality" – a trend also observed in the field of algorithmic fairness. By situating intersectionality within its broader activist and intellectual lineage, highlighting its origins, and acknowledging its epistemic implications, we aim to counterbalance our otherwise necessary reliance on Crenshaw's seminal work.

[2] We use a capital letter to highlight the political dimension of the category Black.

*paradox*, in which the claims of Black women plaintiffs were dismissed on contradictory grounds. In the landmark case DeGraffenreid v. General Motors, five Black women alleged that General Motors had never hired a Black woman prior to 1964, despite employing white women and Black men. The court rejected their discrimination claim, asserting that it did not constitute race discrimination, sex discrimination, or either category independently, but rather an impermissible "combination of both" [2]. Consequently, Black women's experiences, multiply marginalized in terms of gender *and* race, were deemed *too specific* to qualify as race or gender discrimination alone, since not all women or all Black people had suffered the same harm. At the same time, the courts found their claims *not specific enough,* refusing to recognize "Black women" as a "special class to be protected from discrimination." [2] This rendered intersecting experiences "partial, unrecognizable, something apart from standard claims of race discrimination or gender discrimination." [5] Crenshaw's metaphor of a traffic *intersection* where discrimination operates through multiple, intersecting pathways resonated deeply with feminist scholars and has profoundly influenced feminist research since then. It spurred interdisciplinary engagement, leading to the global adaptation of intersectionality across feminist theory, political science, critical legal studies, social movements, public policy, and international human rights [6]. Other axes of oppression were incorporated, including those on migration status, religion, and disability [7, 8], reinforcing its relevance in analyzing how systemic inequalities persist across different social and institutional contexts.

The rise of data-driven technologies, particularly Artificial Intelligence (AI)[3], has significantly altered societal discussions on equity, representation, and discrimination. Often perceived as neutral and objective, invoking an "aura of truth, objectivity, and accuracy," [9] algorithmic systems frequently encode and amplify historical biases rooted in longstanding global and local power structures [10, 11]. Cases of unfairness, sexism, and racism in high-stakes contexts such as credit scoring [12], criminal sentencing [13, 14], and hiring [15, 16], prompted the field of algorithmic fairness[4] to address such issues. Most of the efforts within that field are based on numerical methods, including statistical fairness metrics that measure outcome parity across demographic groups [17–19], individual-based approaches that evaluate fairness by comparing outcomes for similar individuals [20], and causal approaches that attempt to minimize the direct influence of demographic attributes on outcomes [21]. While these approaches contribute to fairness efforts, they face significant criticism: they all address diverse, partly contradictory conceptions of fairness or justice [22] or do not rely on moral foundations at all [23], rely on rigid similarity measures, tend to address symptoms of discrimination rather than its structural roots [24], and depend on predefined structural choices that often overlook procedural [25] and relational [26] dimensions of fairness.

The point of critique that we take up in this paper is that algorithmic fairness methods – similar to the justice system's critique sketched above – predominantly adopt a single-axis framework on discrimination, analyzing fairness through *isolated* attributes such as race or gender. Limited efforts to account for *intersectional biases* in algorithmic systems have taken the form of auditing and adjusting for subgroup disparities (e.g., gender × race), as we will elaborate upon in Section 3.1. As we will see, they computationally undermine what (Black) feminist activists have been fighting for in the last decades: the acknowledgment of discrimination taking place at the system level, the recognition of diverse lived experiences, and the importance of representation of marginalized voices. Intersectionality nowadays can roughly be summarized as examining how multiple social categories, such as race, gender, class, age, or disability, interact to shape lived experiences and how these experiences position people differently within the world, especially within systems of power embedded in specific social, cultural, and political contexts [4] – without being a single, fixed concept but rather a plurality of perspectives, discourses, and debates around its key tenets. As Cho et al. [5] assert, "what makes an analysis intersectional is not its use of the term 'intersectionality,'" but rather the adoption of an intersectional way of thinking about power, sameness, and difference. In this context, historical examples across disciplines suggest that an intersectional perspective can significantly *enhance* efforts towards social justice, not only for those multiply marginalized, but for all people. Reducing intersectional algorithmic fairness efforts to subgroup analyses not only fails to acknowledge intersectionality's complexity and thus catapults discourse back to 1989, but also misses the opportunity to use computational power to enhance social justice efforts actively.

[3] Whenever referring to AI systems in this paper, we adopt the technical interpretation of AI as machine learning systems, while acknowledging the socio-technical nature of such systems. For brevity, we use the term algorithmic systems where appropriate.

[4] While the technical field is called ‚algorithmic fairness', discussions also include justice discourses without being explicitly distinguished. When we talk about algorithmic fairness, we refer to the discourses within this field.

While the reduction of intersectional approaches in algorithmic fairness to formal metrics has been the subject of critique by several scholars (e.g., [27–29], see Section 3.3), and guidance on how AI research might engage with intersectional feminism has been proposed [30], there remains a lack of concrete, actionable recommendations that are grounded in a conceptually robust framework of intersectional feminism. We propose a complementary approach by extending Green's [24] distinction between formal and substantive algorithmic fairness to the intersectional domain. Our goal is to map which parts of current work are formal versus substantive, and to provide AI practitioners with concrete guidance for shifting from formal fairness checks to more context-sensitive, substantively informed practices. Therefore, we ask:

*How can the inherently complex and nuanced nature of intersectionality be substantially integrated into algorithmic fairness research and practice?*

To do so, we first draw on intersectional feminist perspectives to illuminate the complex and nuanced nature of intersectionality in Section 2. In Section 3, we first outline the current state of the art in computational approaches to intersectional algorithmic fairness (Section 3.1), which we summarize as *Formal Intersectional Algorithmic Fairness*. After sketching some inherent computational limitations (Section 3.2), we synthesize critical literature on these computational approaches (Section 3.3), which will serve as the baseline for our substantive approach. In Section 4, we propose *Substantive Intersectional Algorithmic Fairness,* comprising ten desiderata for computational efforts that are both actionable and faithful to intersectionality's structural insights, thereby avoiding the oversimplifications that perpetuate algorithmic harms [31, 32]. We discuss challenges and implications in Section 5.

We emphasize that our contribution is not a unified operational framework, nor do we claim to resolve all structural sources of inequity. Instead, we identify conceptual, epistemic, and methodological requirements necessary for evaluating and guiding algorithmic systems in context. These requirements function as orientation points for practitioners and scholars, clarifying how intersectional theory can reshape what counts as a fairness-relevant consideration without reducing intersectionality to a checklist or formal criterion.

**Authors Positionality**

This paper engages with fairness and justice – deeply normative concepts – and acknowledges that all knowledge is "socially constructed and transmitted, legitimated, and reproduced," both shaping and being shaped by power relations [33]. Intersectionality, in particular, requires a critical examination of the conditions that make its knowledge claims intelligible [33]. Recognizing this, we situate ourselves explicitly within the discourses we engage. We are scholars from mathematics, software engineering, ethics, social sciences, and political science (gender studies). All authors are trained in gender studies, and all but one in computational methods. Based in Germany and the Netherlands, we aim to position our work within a global context. As white scholars, particularly three women and one man, at different career stages, we acknowledge that our perspectives are shaped by our locations within European academia and the privileges and limitations this entails. Grounded in a social justice ethos informed by decolonial thought, we prioritize equity over equality and advocate reparative approaches to historical injustices. In developing *Substantive Intersectional Algorithmic Fairness*, we strive for reflexivity and rigor, recognizing that transparency about our positionality is essential to critically engage with the epistemological foundations of our work.

## 2 Theoretical Foundations of Feminist Perspectives on Intersectionality

Intersectionality has a long, diverse, and often contested history within the social sciences and humanities. Consequently, there is neither a fixed definition nor is it a monolithic theory, but rather an evolving discourse on the complexity and fluidity of lived experiences and oppressive power structures [33]. In this section, we trace these intellectual lineages, clarify conceptual foundations, and introduce several core commitments that inform our feminist approach to Substantive Intersectional Algorithmic Fairness.

As we have touched upon, intersectionality historically emerged from *social justice movements* by people who were "deeply concerned by the forms of *social inequality* they either experienced themselves or saw around them." [4] The most visible lineage stems from U.S. Black feminist thought [33] in which women, later redefined as women of colour, were organized in grassroots movements and other multiple forms of local community organizing [34]. Their activism emphasized the necessity of viewing lived experiences of Black women confronting racism, sexism, classism, and heteronormativity simultaneously as intertwined rather than discrete or hierarchical phenomena [35]. Movements similar to U.S. activism emerged globally, often independently. For instance, in the United Kingdom, Stuart Hall's cultural studies explored the entanglements of class, nation, race, and ethnicity in shaping immigrant experiences and British multiculturalism [36]. In continental Europe, scholars such as Nira Yuval-Davis and Floya Anthias analyzed multiple dimensions of oppression within specific socio-political contexts [37]. All of this activism laid the groundwork for intersectionality as transformative praxis from the ground up, aimed at dismantling systemic injustice, fighting for social justice, and collective liberation [38, 39].

When Crenshaw [2, 3] described in her articles how women of colour experience male violence as a product of intersecting racism and sexism, yet were marginalized within both feminist and anti-racist spaces and legal remedies, she found a powerful way of *articulating* what many grassroots movements and social activists had long been fighting for. As a consequence, Crenshaw's metaphor became the face of intersectionality. However, when entering academia and policy discourses, intersectionality soon became a framework for merely *describing* and *diagnosing* social inequalities [1]. As a consequence, activist perspectives were frequently depolitized, stripped of their critical edge, and reframed to align with dominant epistemic norms, such as scholarly objectivity and methodological individualism [4]. Through mechanisms such as patterns of citation, themes of journal articles, invitations to deliver keynotes, and the composition of panels at academic conferences, the narrative about intersectionality were controlled, while alternative perspectives were ignored and suppressed [1]. At the same time, intersectionality did indeed become a powerful analytical tool to map which causal mechanisms produce inequalities, how inequalities are structured, and how power operates [5, 33]. Since hierarchies are neither fixed nor predetermined, this remains an ongoing task [40–42]. Nowadays, intersectionality is widely regarded as both: a critical praxis and an analytical tool, and Collins und Bilge [4] highlight the "interconnections between the two." Intersectionality as an *analytical tool* maps inequalities as *what they are*, not necessarily prescribing political action and transformative work. Intersectionality as *critical praxis* is about *what should be*, thus fostering organized action by guiding social movements, policy-making, and institutional reform [5, 33, 42].

In both views, there are recurrent topics which Collins und Bilge [4] summarize in six core themes: apart from social justice and social inequality, intersectionality is engaged with relationality, power dynamics, social context, and complexity.

Across its diverse genealogies, intersectionality rejects reductionist frameworks that treat social categories[5] as isolated, mutually exclusive, or inherently oppositional, such as simplistic binaries of Black versus white, or old versus young [44]. Instead, it examines how these categories co-construct one another, emphasizing the dynamic, interdependent, and context-specific nature of social identification [42, 45, 46]. This refusal of *either/or binary thinking* in favor of *both/and thinking* foregrounds the *relational* character of social differentiation. Relational theories conceptualize inequality as a social and political relation of domination and subordination rather than merely a distributive imbalance [47]. In this view, unfair distributions are *symptoms* of social injustices, and, as Young [48] argues, "the concepts of domination and oppression, rather than the concept of distribution, should be the starting point for a conception social justice." Relational justice thus requires reciprocity and mutual respect, ensuring that no one perceives themselves as inherently superior or inferior to others, whether in interpersonal interactions [49] or institutional structures [48]. That is why intersectionality rejects the notion that any form of oppression is more fundamental or more worthy of redress than another [50]. Enacted through dialogue, interaction, and other forms of engagement, the relational perspective reframes social inequality not as a "race-only" or "gender-only" problem, but as the product of interactions among multiple, co-constitutive categories [4].

This relational lens leads directly to intersectionality's core concern with *power dynamics* embedded in institutions, ideologies, and social norms as a central element of systems of oppression and the production of social inequalities. Cudd [51] distinguishes between objective oppression, referring to material constraints on action and well-being, and subjective oppression, referring to the internalization of oppressive norms and beliefs. Thus, power is embedded in the relations that constitute social life [4]. This means that intersectionality is concerned with the -isms and their mutual influence: racism, classism, and sexism on a micro-level [46], and neoliberalism, nationalism, and capitalism on a global level [4]. These systems operate simultaneously across multiple domains: structural (e.g., laws, policies), disciplinary (e.g., bureaucratic practices), cultural (e.g., dominant narratives), and interpersonal (e.g., everyday interactions) levels [4, 40]. Acknowledging power relations challenges the notion of detached 'objectivity': positionality matters, and marginalized standpoints provide an essential, though often undervalued, lens [42, 45, 46]. Thus, intersectionality requires uncovering the epistemological assumptions that shape our understanding of social phenomena [33].

Importantly, intersecting power relations and their role in shaping social inequalities must be understood in relation to their specific social, historical, and cultural *contexts* [4]. Such contexts profoundly influence how individuals perceive themselves, interpret the world, and act within it, while also shaping the construction and relative *salience* of social categories and their interconnections. For instance, Abdellatif [52] recalls that her gender identity was most salient in religious settings while living in Egypt, whereas in the United Kingdom her experiences foregrounded the intersections of race and precarity. Even within ostensibly shared environments, divergent social positions can yield fundamentally different perspectives and lived experiences [4]. A key insight from these diverse intellectual traditions is that the salience of race, class, gender, or other axes of identity is not fixed but varies across time and place, reflecting the shifting configurations of power in which they are embedded. Intersectionality, therefore, demands attentiveness to contextual specificity and a rejection of universalist assumptions that treat social categories as stable, homogeneous, or universally experienced.

[5] Social categories can emerge through processes of social categorization imposed from outside, become constitutive of individual and collective identity, or be institutionalized in law and policy as protected characteristics, see for example [43]. All three purposes can be relevant in the context of algorithmic systems, so we ask for careful consideration of the nuances while reading this article.

Interestingly, Collins und Bilge [4] emphasize that *complexity* is not only what intersectionality seeks to illuminate but is also intrinsic to intersectionality itself. Intersectionality is designed to grapple with the interwoven nature of social categories, lived experiences, and domains of discrimination, all of which are embedded within broader dynamics of inequality, power, relationality, and context – elements that are themselves mutually constitutive. Thus, it is a core characterization of intersectionality to not being able to provide a "crisp instruction manual for applying intersectionality to various fields of practice." [4] Instead, the commitment to complexity, rather than chasing quick fixes, is essential for both theoretical rigor and practical applications in the light of intersectionality.

As we have seen, intersectionality today consequently lacks a single, universally agreed-upon definition. Its adaptability enables application across a broad range of social, cultural, and disciplinary settings, and, as Davis [53] notes, "it is precisely the concept's alleged weaknesses – its ambiguity and open-endedness – that were the secrets to its success and, more generally, make it a good feminist theory." Yet this conceptual openness, while a source of strength, also resists any fixed or static operationalization. As Collins [33] cautions, "presenting a finished definition of intersectionality that can be used to determine whether a given book, article, law, or practice fits within a preconceived intersectional framework misreads […] intersectionality's complexity." In computational settings, however, this becomes a practical challenge [54]. Said [55] and Knapp [56] describe how social theories can lose their original meaning when transposed from one domain to another, often compromising their depth in the process. Intersectionality faces a similar danger when being adopted in the field of algorithmic fairness, as illustrated in the next section.

# 3 Formal Intersectional Algorithmic Fairness

When Buolamwini und Gebru [57] illustrated in *Gender Shades* that algorithmic systems could *appear* fair when evaluated separately by gender or race, yet yield highly distorted results for overlapping subgroups such as female-white, male-white, female-Black, male-Black[6], they set the ball rolling for some researchers to explore *subgroup-based* fairness assessments instead of the so-far predominant focus on single-axis fairness metrics. These subgroup-based fairness assessments seemed to be inspired by an intersectional lens on algorithmic fairness and were consequently often referred to as *intersectional (algorithmic) fairness*. We will collectively refer to these subgroup-based fairness assessments as *Formal Intersectional Algorithmic Fairness*, aligning our critique with Green's [24] notion of formal algorithmic fairness, which conceptualizes fairness as a "fair contest" where all individuals are evaluated by the same standard, based solely on their characteristics at the moment of decision-making [24]. While formal (intersectional) algorithmic fairness is invaluable for shaping computational fairness debates, it is also characterized by a high degree of abstraction and rigidity. In this section, we will first review some of these subgroup-based, formal approaches to intersectional algorithmic fairness (Section 3.1). We will then outline some of their inherent computational limitations (Section 3.2) before highlighting the conceptual, methodological, and normative concerns that some scholars have raised, particularly regarding their treatment of systemic factors of discrimination (Section 3.3). We distill six main areas of critique.

[6] This phenomenon was later called *fairness gerrymandering* [58]. It can easily be imagined in a classifier that outputs 100% positive results for male-white and female-Black, but 0% positive results for female-white and male-Black. It would yield 50% positive results for both genders and both races.

## 3.1 Formal Approaches to Intersectional Algorithmic Fairness

In a recent survey, Gohar und Cheng [59] distinguish between a posteriori and a priori approaches to intersectional algorithmic fairness. *A posteriori metrics* evaluate algorithmic fairness after outcomes are produced, seeking parity across subgroups. Kearns et al. [58], for instance, extend classical criteria such as equal opportunity and statistical parity to a theoretically unbounded set of subgroups. These subgroups, defined prior to testing, may include not only demographic and legally protected attributes but also any relevant partition of the data. This approach statistically struggles to address small or underrepresented groups while maintaining mathematical guarantees. Relatedly, Foulds et al. [60] adapt the U.S. Equal Employment Opportunity Commission's 80% rule to multi-attribute contexts. This extension incorporates multidimensional subgroups but faces statistically similar challenges due to insufficient data for smaller groups. Ghosh et al. [61] introduce Max-Min Fairness, an approach that maximizes the minimum utility across all subgroups. By uncovering hidden inequalities, this method can benefit small groups (in terms of representation within the data); however, it becomes increasingly limited by data sparsity issues as the number of dimensions increases. By contrast, *a priori metrics* seek to calibrate outcomes in alignment with the demographic characteristics present in training data. Hebert-Johnson et al. [62] propose multicalibration, which requires consistent calibration across all identifiable subgroups under the assumption that subgroups can be sufficiently represented and reliably detected. Related efforts, such as those of Kim et al. [63] and Gopalan et al. [64] extend this calibration principle, while Yona und Rothblum's [65] Probably Approximately Metric-Fairness builds on Dwork et al.'s [20] similarity-based framework, ensuring that similar individuals receive similar outcomes with a bounded error rate. These approaches, however, depend critically on the availability of technically accurate subgroup definitions, adequate subgroup representation in the data, and the assumption that training datasets provide a sufficiently uniform reflection of the underlying population.

## 3.2 Computational Limitations

Beyond the individual shortcomings of these metrics, there are several overarching *computational limitations*. A priori and a posteriori approaches are not simultaneously satisfiable: a priori methods align model outcomes with training data distributions, thereby risking the reinforcement of historical biases (sometimes an exact comparison with the training data is simply not desired). A posteriori methods, by contrast, enforce outcome parity across subgroups, requiring post-hoc adjustments that often introduce a trade-off between predictive accuracy and fairness. This tension forces practitioners into a choice between methods that either replicate existing inequities or compromise model performance. Chen et al. [66] highlight an additional difficulty: in their comparison of eleven fairness-improvement methods across multiple protected groups, they show that optimizing fairness for one subgroup can diminish fairness for subgroups not explicitly addressed, even when overall accuracy remains stable. Similarly, Huang et al. [67] examine subgroup fairness within the context of explainable AI (XAI) and demonstrate how popular methods systematically *underestimate* the importance of subgroup biases. For example, SHAP evaluates the influence of individual features (race, gender, etc.) on predictions by manipulating them independently. Since "being a Black woman" represents the intersection of multiple characteristics, it cannot be treated as a discrete, manipulable feature. This also applies broadly to counterfactual fairness evaluations, making *individual* notions of fairness generally particularly challenging. Another persistent computational challenge in all of these approaches is determining the appropriate number and selection of subgroups. Kearns et al. [58] point to the *exponential* number of ways a population can be divided into partitions. Observing that while it is infeasible to identify a small number of subgroups as the sole focus of fairness interventions, they emphasize that insisting on statistical fairness *for every possible* subgroup risks overfitting and undermining classifier generalization. Thus, from a strictly computational perspective, formal approaches to intersectional fairness face severe limitations.

## 3.3 Conceptual, Methodological, and Normative Concerns

Apart from the computational limitations, some scholars have raised conceptual, methodological, and normative concerns about formal approaches to intersectional algorithmic fairness [27, 28, 68–71]. These critiques converge around three broader key areas, within which we identify six distinct aspects that highlight recurring blind spots and tensions (see Table 1 for an overview). These will provide an essential foundation for our substantive approach presented in Section 4.

**Table 1** *Key Areas and Aspects of Formal Intersectional Algorithmic Fairness*

| *Key Area* | *Aspects* |
|---|---|
| I) Use of Protected Attributes | (1) Reliance on statistically significant subgroups<br>(2) Centering disadvantage over systemic privilege |
| II) Neglect of Social Context and Consequences | (3) Categories are detached from power<br>(4) Ignoring the consequences of algorithmic outputs |
| III) Roles of Researchers and Practitioners | (5) Veneer of fairness through individualist framing<br>(6) Lack of reflexivity and responsibility among researchers and practitioners |

*3.3.1 Reliance on statistically significant subgroups.* Hoffmann [68] and Hampton [69] critique treating categories such as gender or race as fixed, independent variables, which results in an oversimplified "stacking" of oppressions, reducing complex experiences to additive disadvantages. Kong [28] further highlights the dilemma of category selection: deciding which categories to choose leads to a tension between "infinite regress" – where endless subdivisions of categories are created – and the arbitrary selection of the crossover of categories. Proposals to address this, such as focusing only on statistically significant subgroups, risk neglecting subgroups that are underrepresented within the data, mostly already marginalized communities, thus failing to resolve the political nature of deciding what is "relevant." Himmelreich et al. [71] further argue that omitting underrepresented subgroups violates prioritarian principles, since these groups often face the gravest injustices.

*3.3.2 Centering disadvantage over systemic privilege.* Several scholars note that fairness research focuses primarily on mitigating disadvantage while neglecting systemic privilege. Hoffmann [68] observes that privilege is normalized and rendered invisible, which allows dominant norms resulting from, for example, whiteness or maleness, to persist unexamined. Hampton [69] and Davis et al. [70] similarly argue that focusing on disadvantage without interrogating privilege risks reinforcing the very hierarchies that fairness interventions seek to address.

*3.3.3 Categories are detached from power.* Kong [28] critiques fairness metrics for reducing social categories to static labels, thereby divorcing them from histories of oppression such as colonialism, sexism, or racism. Davis et al. [70] and Hoffmann [68] similarly point out that this "flat" treatment [in reference to 72] of categories erases systemic inequality by ignoring how identities are shaped through power and history. Ovalle et al. [27] find that research papers that even initially acknowledge power often operationalize intersectionality in purely categorical terms, effectively sidelining its critical dimension. The neglect of social context and power structures in algorithmic fairness debates is closely linked to discussions of distributive conceptions of justice. Hoffmann [68] criticizes that algorithmic fairness research only considers discrimination to be *relevant* when it can be *visible* in particular unequal *distributions* of outcomes. A distributive account in this context treats fairness as a question of who gets what, rather than asking why certain groups are systematically disadvantaged or privileged. Building on this, Kong (2022) highlights longstanding critiques of the distributive paradigm, most notably by Young [48] who contends that distributive approaches obscure the institutional and structural conditions that produce inequality. Similarly, Davis et al. [70] emphasize that algorithmic fairness discourse often presumes a meritocratic social order, seeking to correct disparities without interrogating the structures that produce them in the first place.

*3.3.4 Ignoring the consequences of algorithmic outputs.* Scholars also warn against evaluating fairness without considering the lived consequences of algorithmic decisions [68, 70]. Equal error rates may mask vastly different effects: for instance, the denial of a loan may be inconvenient for a privileged applicant but devastating for someone facing multiple dimensions of oppression [68]. Ovalle et al. [27] further note that some scholarly work assumes improved subgroup fairness equates to social justice.

*3.3.5 Veneer of fairness through individualist framing.* Hoffmann [68] criticizes the dominant conceptualization of discrimination in many algorithmic fairness debates as being overly individualized and behaviorist in nature. In particular, she criticizes the narrow, individualist view of discrimination rooted in what Freeman [73] calls the perpetrator perspective, which is a view in which discriminatory harms are reduced to isolated acts of individual misconduct. This framing leads to a narrow *causal* understanding of discrimination: discrimination is seen as a matter of bad actors who deviate from a norm of impartiality, rather than as a product of deeply embedded social structures that continuously produce inequality. This leads to the goal of neutralizing inappropriate conduct on the part of individuals instead of resolving all contributing conditions [73]. Hoffmann [68] warns that if computational approaches focus on neutralizing improper uses of protected attributes in computational systems, the conditions that enable discrimination remain untouched, and fairness becomes a formalist exercise detached from material outcomes. Thus, the emphasis on individual acts and categorical parity may offer a *veneer of fairness*, while diverting attention away from institutional responsibility and systemic reform. This veneer of fairness is taken up by Himmelreich et al. [71]. They argue that some computational approaches encourage "perverse" behaviour, i.e., in terms of discouraging data collection and allowing "gaming" (e.g., making intentionally incorrect predictions to appear fair). Some metrics, for instance, might punish adding data for an underrepresented group since it might tighten fairness constraints, making a model *seem* less fair.

*3.3.6 Lack of reflexivity and responsibility among researchers.* Finally, even though reasons for systemic discrimination might not be transferred to individuals, Ovalle et al. [27] critique the insufficient reflexivity among researchers and practitioners. According to them, many researchers frame power dynamics as inherent properties of the technical system, absolving themselves of responsibility for reproducing inequalities. However, the "notion that a system itself exerts power is technodeterministic, i.e., it reifies the idea that systems, and not their creators, are responsible for reproducing inequalities.” [27] In addition to that, when social context is mentioned in research papers, it is frequently relegated to a “limitation” rather than recognized as constitutive of the research process itself.

Taken together, these six aspects demonstrate how formal approaches extend beyond single-axis fairness models but remain constrained by their reliance on fixed categories, abstraction from social context, and insufficient reflexivity (key areas). In Section 4, we will now articulate how adhering to intersectionality in its theoretical origins can reorient algorithmic fairness research and practice.

# 4 Substantive Intersectional Algorithmic Fairness

Building on the theoretical foundations of intersectionality (Section 2) and the critique of Formal Intersectional Algorithmic Fairness (Section 3), we synthesize recommendations to advance a substantive approach to Intersectional Algorithmic Fairness. We present these recommendations as a set of desiderata, a term derived from the Latin desideratum, meaning “something considered essential.” Intended to cultivate a culture of conscientious practice, they should be understood as an invitation for those seeking to approach intersectionality in a meaningful way. At the same time, given the conceptual and practical complexity of intersectionality, they constitute minimal requirements rather than an exhaustive checklist. Further, they address both intersectionality as an analytical tool and a critical praxis, while preserving the interconnectedness between the two. Importantly, these dimensions are neither discrete nor independent but interconnected, each emphasizing distinct yet complementary aspects of intersectionality.

**Table 2** *Desiderata for Substantive Intersectional Algorithmic Fairness*

|  |  |  |
|---|---|---|
| *I) Recognizing Basic Epistemological Assumptions* | 1 | Interrogating assumed neutrality in decision processes. |
| | 2 | Making positionality explicit. |
| | 3 | Being conceptually precise, explicitly addressing oppression, and moving beyond the language of ‘ intersectional algorithmic bias’. |
| *II) Overcoming the Narrow Focus on Protected Subgroups* | 4 | Questioning the meaning of social categories. |
| | 5 | Not weighing or ordering oppression. |
| *III) Overcoming the Lack of Attention to Socio-Technical Systems* | 6 | Mapping power and domination structures. |
| | 7 | Acknowledging that small actions can have a significant impact – and a distinct impact for distinct groups. |
| | 8 | Aligning purpose with context and impact of actions. |
| | 9 | Explicitly considering privileges and not only disadvantages. |
| *IV) Acknowledging Transformative Potential* | 10 | Recognizing algorithmic systems’ opportunity for intersectionality beyond critique. |

## 4.1 Recognizing Basic Epistemological Assumptions

As Crenshaw [2] discussed in her article, legal fairness judgments often relied on the assumption that categories like race or gender interfere with otherwise “fair” or “neutral” decisions. Similarly, when Formal Intersectional Algorithmic Fairness addresses discriminatory patterns by utilizing subgroup fairness metrics, this implicitly assumes that algorithmic processes are fair or objective *as long as* explicit discriminatory patterns are quantitatively revealed and sufficiently explained [23, 30]. This implicit claim of neutrality obscures the social and historical conditions under which the knowledge that has influenced the design of the algorithmic system at hand, as well as the conditions under which it operates, has been materialized. In other words, algorithmic systems' design is deeply influenced by developers' epistemic backgrounds, is grounded in epistemic injustices in society, and functions as epistemic technology itself. Dotson [74, 75] conceptualizes *epistemic oppression* as the systematic exclusion of marginalized groups from participation in knowledge production, thus suppressing marginalized epistemic agency while elevating dominant perspectives. A substantive approach to Intersectional Algorithmic Fairness requires an intense engagement with epistemological claims, since epistemic oppression is a core defining feature of intersecting systems of power [1].

In desideratum 1, we thus first discuss how epistemic injustices within society influence algorithmic decision-making, then elaborate on how algorithmic systems reinforce them, and, in desideratum 2, turn to developers' and users' epistemic backgrounds. Desideratum 3 argues that clear, precise language and a realistic assessment of algorithmic systems' potential are prerequisites for a substantive approach to Intersectional Algorithmic Fairness.

**Desideratum 1.** *Interrogating assumed neutrality in decision processes.*

Since algorithmic systems are trained on historical data and operate on socially produced inputs, such as CVs in hiring contexts, the epistemic injustices embedded *within society* inevitably manifest within the algorithmic pipeline. Interrogating assumed neutrality, therefore, requires examining how these injustices structure what is taken as evidence, whose experiences become legible to the system, and which forms of knowledge are excluded. Rather than treating harmful patterns as deviations from an otherwise neutral process, a substantive intersectional approach demands engaging with and understanding these epistemic injustices to counteract them.

For instance, t*estimonial injustice* occurs when individuals receive less credibility due to prejudices about their social identity [76]. In a hiring context, some Black women might refrain from applying because they have internalized the expectation that their competence will be doubted – a subjective reaction to objective oppression[7]. When they do speak to their achievements, they might be perceived as *aggressive* or *self-promoting*, a result of the intersection of racialized and gendered stereotypes. *Hermeneutical injustice* occurs when individuals lack the shared conceptual resources needed to interpret or express their own experiences [76]. Black women who have finally applied may have earned resilience, conflict-resolution skills, and social intelligence by implicitly navigating systemic racism or acting as informal caretakers, yet may not fully recognize these capacities as skills or present them as such in professional contexts. *Contributory injustice*, similarly, refers to situations in which marginalized groups possess relevant epistemic resources, but dominant institutions fail to acknowledge or integrate them [74]. Hiring firms may therefore overlook these earned capacities – not because they lack value, but because dominant frameworks are unable or unwilling to recognize them. While these forms of injustice affect many marginalized groups, they disproportionately impact those who are intersectionally marginalized, as we have seen in Crenshaw's [2] elaboration on how Black women's experiences have historically been framed as simultaneously "too specific" and "too general" to fit dominant interpretive schemas. When such experiences are unrecognized in society, they remain invisible within conventional hiring narratives and, by extension, within the algorithmic systems built upon them.

[7] To be more precise, this would be an example of indirect oppression of choice; we refer the interested reader to [51].

Two consequences follow. First, when Formal Intersectional Algorithmic Fairness evaluates only downstream disparities without interrogating the epistemic conditions that generate them, it cannot fully assess whether an algorithmic system is fair.

Second, algorithmic systems increasingly function as epistemic technologies that *reinforce* existing inequities [77, 78]. For instance, Hardalupas [79] introduces the term *epistemic calcification* to describe how AI systems "harden" dominant epistemic resources, thereby contributing to contributory injustice. Several mechanisms drive this process [79]: First, many algorithmic systems rely on supervised learning, which uses labels and a structure imposed by developers and are thus shaped by those defining the labels [80] – a fact that we will return to in desideratum 2. Second, algorithmic systems impose a certain way of solving a problem. For example, solely relying on visual signals to detect dermatological conditions on darker skin, or omitting trans and non-binary categories from COVID-19 surveillance data [81], inevitably influences the problem's solution. Third, the inherent promise of scalability in AI presupposes user homogeneity [82] and thus renders particularly intersectionally marginalized experiences invisible by design. This factor is taken up by Miragoli [83] who describe AI systems as gatekeepers of epistemic resources. They explain that *epistemically conformist behaviour* (treat statistically dominant patterns as authoritative, while minority experiences are noise, aberrations, or insufficiently representative) is responsible for the marginalization of minoritarian perspectives. This produces forms of epistemic oppression where marginalized groups' knowledge is not only excluded but actively overwritten. Thus, the drive for predictive efficiency incentivizes ignoring multiply marginalized experiences. As a consequence, epistemic conformism contributes to structural injustice by legitimizing majority-based patterns as neutral facts. Lastly, automation bias, which describes the overreliance on automated systems despite contrary evidence, further entrenches the dominance of the "knowledge" of AI systems [79].

Thus, precisely the field's motivating goals continue to reinforce epistemic calcification by naturalizing dominant ways of knowing. Working toward intersectional justice would require confronting and transforming the epistemic infrastructures that produce contributory injustice. However, recognizing that decision processes – including algorithmic ones – are never epistemically neutral is a prerequisite and thus a minimal desideratum. While sounding obvious, research continuously demonstrates that recognizing the non-neutrality of technology is far from common in technical disciplines: engineers often treat systems as value-free instruments [84], and abstraction practices in algorithmic design routinely obscure structural power relations [31]. Scholars such as Benjamin [85], Noble [86], or Birhane [87] further show that neutrality claims themselves serve to legitimize and reproduce racialized and gendered hierarchies.

**Desideratum 2.** *Making positionality explicit.*

As we have seen, algorithmic systems increasingly function as epistemic technologies themselves [77, 78]. Algorithmic systems inherently embody assumptions that are made within the learning process about what constitutes valid information, which problems are worth solving, and which solutions legitimate [17, 31]. This desideratum turns to the positionality of the agents, institutions, and epistemic standpoints that shape those processes. Without explicit attention to positionality, the questions "Who is represented, and how?" and "Who is empowered, and who is rendered invisible?" cannot be answered with sufficient rigor. A substantive intersectional approach to algorithmic fairness requires that the situated standpoint from which knowledge is produced is made visible and systematically integrated throughout the development and evaluation of algorithmic systems.

Standpoint epistemology, rooted in feminist and critical race theory, holds that social position shapes what individuals and institutions perceive as problems, what they accept as valid evidence, and which interpretations they deem meaningful [40, 88, 89]. Given the demographic and epistemic homogeneity of the machine learning field, dominated by WEIRD (Western, Educated, Industrialized, Rich, Democratic) contexts [90], algorithmic systems inherently reflect dominant worldviews, materializing in normative choices about data collection, model objectives, ground-truth labels, and deployment contexts. This reproduces epistemic oppression through both the demographic composition of practitioners as well as the, probably resulting, *epistemological orientation* of the discipline, which is often governed by technical rationality and techno-solutionism: the belief that complex social problems can be solved through technical means alone [11, 91–93]. As Collins [1] warns, the *refusal to acknowledge* structural oppression in the context of intersectionality suppresses the conceptual tools necessary to *analyze and confront* them, producing what Nguyen [94] describes as an epistemic echo chamber in which agents are not merely uninformed but are taught to distrust external perspectives. Coeckelbergh [78] argues that algorithmic systems increasingly mediate the belief formation of those who use them, making it even more difficult for structurally, intersectionally disadvantaged individuals to exercise epistemic agency and revise their understanding of the world.

Making this positionality explicit means institutionalizing epistemic reflexivity throughout the algorithmic lifecycle [e.g., 95]. This means, first of all, recognizing one's own position within epistemic practice, then making it publicly accountable, and, lastly, shifting the epistemic center of algorithm design. Current scholarship offers several concrete, actionable strategies to do so.

The *first* step of making positionality explicit requires a critical interrogation of the normative assumptions in data and design – thus a reflection of one's own positioning and the normativity behind one's actions, as well as the positioning and representation of those subject to the algorithmic system. Who is represented, and how? Who is empowered, and who is rendered invisible? What do social categories such as 'race' or 'gender' signify in the given context? For example, can a Black woman be adequately represented under the singular category "Black;" does the term "woman" primarily refer to white women? Leavy et al. [95], for example, propose a framework that requires developers to examine the perspectives embedded in data, make the reflexive nature of knowledge production explicit, analyze the theoretical assumptions underlying datasets, and include subjugated or emergent knowledge systems. Their emphasis on an "active anti-racist stance" aligns with intersectionality's commitment to social justice. QuantCrit, introduced by Gillborn et al. [96], is a framework for making epistemic standpoints in statistical analysis visible and aligning them with anti-racist principles. The *second* step is then to make positionality publicly accountable. One concrete mechanism is integrating positionality statements into research publications, model cards, or software documentation. These statements disclose the epistemic assumptions behind methodological decisions and locate the designers within the socio-political structures that shape their work. Such transparency enables evaluators, users, and affected communities to assess the appropriateness of an algorithmic system for the intended task and context. *Third*, the epistemic center of algorithmic design must shift. Standpoint theory argues that individuals at the intersections of structural oppression often possess distinctive epistemic insights into systems of power, precisely because they are required to confront their effects in everyday life [40, 88, 89][8]. These insights do not derive from identity alone, but from critical reflection and collective experience. That is why increasing diversity in the tech sector is not merely a question of representation, but a necessary condition for epistemic justice. Participatory frameworks such as *Design Justice* [98] operationalize this shift by explicitly challenging the dominant practice of imposing the designer's standpoint onto technological systems and instead centering those most affected by algorithmic systems in key design decisions, redistributing epistemic authority beyond dominant groups, and legitimizing alternative ways of knowing within the design process.

[8] For the difference between epistemic advantage and epistemic privilege, we refer the interested reader to [97].

In summary, making positionality explicit is a minimal requirement for a substantive approach to intersectional algorithmic fairness that engages with and makes explicit the underlying epistemological assumptions within algorithmic systems.

**Desideratum 3.** *Being conceptually precise, explicitly addressing oppression, and moving beyond the language of ' intersectional algorithmic bias'.*

This desideratum addresses the epistemic and political consequences of euphemistic language in Formal Intersectional Algorithmic Fairness (see also Section 3.3.5). Phrases like "intersectional algorithmic bias" suggest engagement with intersectionality, yet, as Collins [33] cautions, work invoking intersectionality without interrogating its own epistemological assumptions risks reproducing the very social inequalities it aims to address. This mirrors the depoliticization of intersectionality's emancipatory origins when absorbed into academia (as discussed in Section 2), where foundational concepts such as standpoint epistemology were sidelined as subjective or essentialist despite their grounding in reflexive, relational knowledge production. This intellectual sanitization is reproduced in Formal Intersectional Algorithmic Fairness, where intersectionality is invoked rhetorically but reduced to counting demographic subgroups within statistical frameworks. As Collins und Bilge [4] argue, such interpretations recast intersectionality as a theory of multiple identities, stripping it of its structural and epistemic critique of power. The shift away from social justice commitments "passes virtually unnoticed" [1], enabling actors "to install a new narrative of intersectionality that privileges academic norms of objectivity and truth," thereby reinforcing their own perspectives while suppressing others. This selective uptake constitutes epistemic violence; the silencing, distortion, or erasure of marginalized knowledge systems that do not conform to institutionalized standards of intelligibility [75]. A central mechanism is linguistic: politically charged terms such as "oppression," "racism," or "colonialism" are replaced with technocratic labels like "bias" or "distributional disparity." The popular term "algorithmic bias" frames harm as an unintended, correctable deviation from neutrality rather than as the result of historically entrenched systems of domination and thus waters down the discussion [69]. Resistance to terms like "oppression" signals an unwillingness to confront the depth of structural inequality [99], a tendency equally visible in wider discourses on intelligence and technology [100].

A substantive approach must reject sanitized terminologies and euphemisms and adopt a conceptually precise, politically accountable language. As Tacheva und Ramasubramanian [101] argue, this requires cultivating a critical attitude capable of "detecting, acknowledging, resisting, and, when necessary, refusing the 'false narratives of progress'" that pervade algorithmic discourse. Naming oppressive systems strengthens design justice [102]. This entails epistemic repoliticization: resisting the reduction of intersectionality to statistical subgroup analysis and reaffirming its roots as a critique of interlocking systems of oppression. In practice, this means explicitly naming oppression (as well as a conceptualization of its precise forms, see for example Cudd [51]), addressing racism, sexism, or colonialism rather than abstracting them into "bias", and shifting epistemic authority from dominant academic or technical actors to those whose lived experiences reveal the workings of structural harm. In short, a substantive approach to intersectional algorithmic fairness demands conceptual and political clarity:

algorithmic harms are not merely the result of "biased data," but symptoms of historically situated, intersecting systems of domination. Naming these systems is a prerequisite for confronting and dismantling them.

## 4.2 Overcoming the Narrow Focus on Protected Subgroups

Having sketched that being transparent about algorithmic systems' epistemological assumptions is a prerequisite for a substantive approach, this section will center on subgroups. In particular, we will discuss the meaning of social categories in desideratum 4 and the hierarchical ordering in desideratum 5.

**Desideratum 4.** *Questioning the meaning of social categories.*

Formal Intersectional Algorithmic Fairness relies on socially constructed categories such as race, gender, or disability to determine whether discrimination is present (see also Section 3.3.1)). Protecting these categories has been instrumental for providing the legal and conceptual foundation for exposing institutionalized forms of oppression through statistical evidence and legal claims [103]. However, while these categories remain essential for identifying and addressing systematic disparities, we observe the tendency to misinterpret them, stripping them of their historical and political contingency. Datasets do not merely record categories; they actively construct and enforce them. The moment a dataset encodes "race" as a fixed label (e.g., Black, White, Asian) or "gender" as a binary (male, female), it operationalizes particular social assumptions while erasing others. For instance, a dataset that records "disability" as a single binary field may flatten the lived differences between, say, neurodivergence and mobility impairment, erasing distinct forms of support needs. Similarly, labeling someone simply as "Black" in contrast to "white" ignores how racialization is distinct between Hispanics, Asians, or Africans, how this racialization changes upon context, and how race intersects with gender, class, or migration history. Measuring fairness only through comparisons of algorithmic outcomes across such predefined groups reduces dynamic, relational identities to static, mutually exclusive data points [68, 69], imposing a *standardized lens on group members*. This mirrors Crenshaw's [2] critique of anti-discrimination law: the legal system, by insisting on discrete identity frames, compels individuals to articulate their experiences within singular, *recognized* categories. Under this logic, a Black woman who is denied a loan because of the compounded effect of race and gender bias may find that neither "race" discrimination metrics nor "gender" discrimination metrics capture her specific harm. Moreover, this framing recognizes discrimination only when it aligns with existing statistical categories and is legible as a measurable disparity. Such technocratic narrowing transforms people into "objects of academic knowledge," stripping them of epistemic agency [1]. Crenshaw [2] notes that "the continued insistence that Black women's demands and needs be filtered through categorical analyses...guarantees that their needs will seldom be addressed." While this does not necessarily reflect deliberate discriminatory intent, it describes "an uncritical and disturbing acceptance of dominant ways of thinking about discrimination." [2] Even when subgroups are formed, these remain tethered to existing categorical boundaries. This produces two problems. First, a combinatorial explosion; the number of intersections multiplies beyond practical measurement, incentivizing selective attention to only the "largest" or most "statistically significant" groups (see also Section 3.2). In addition to that, it creates hierarchies of visibility as certain forms of discrimination (e.g., cisgender women of a dominant racial group) may be more easily detected and addressed than others (e.g., non-binary migrants with disabilities). We will discuss the aspect [104]of hierarchization further in desideratum 5.

A substantive approach to intersectional algorithmic fairness must move beyond the uncritical deployment of predefined categories, without abandoning them altogether. Rather, they should be redefined: categories should be treated as historically contingent, politically constructed, and analytically provisional. Rather than as neutral descriptors, a substantive approach reclaims their critical function as tools that emerged from collective struggles to expose injustice and claim rights [105, 106]. This requires interrogating how categories are defined in datasets, by whom, for what purposes, and with which perspectives excluded, as well as documenting their provenance, contextual meanings, and potential harms of misclassification or omission. Practices such as "datasheets for datasets" [107], "data statements" [108], or "model cards" [109] can support this reflexive documentation, but their use should be embedded within participatory design processes that give affected communities the authority to steer, contest, or veto data and model use. Participatory design methods shift epistemic authority toward those most affected [25, 98, 110], and allow self-identification of categories to map evolving across contexts and over time in recognition of their political and historical contingency. We will discuss these aspects in more detail in the next desideratum.

**Desideratum 5.** *Not weighing or ordering oppression.*

As we have touched upon in the previous desideratum, Formal Intersectional Algorithmic Fairness introduces hierarchies of oppression by choosing which subcategories are *worthy* of investigation. Formal Algorithmic Fairness research tends to focus on the categories for which large, clean, and labeled datasets are available, making statistical tractability and not contextual relevance the primary determinant of which social groups receive fairness interventions (see Section 3.3.1)). This bias is deeply rooted in the politics of dataset construction: Hanna et al. [106] demonstrate how demographic categories in datasets are standardized through bureaucratic processes such as census classifications, medical coding schemes, or platform survey instruments, thereby privileging certain demographics. These classificatory infrastructures constrain algorithmic fairness metrics to administratively convenient categories. As a result, already visible populations benefit from statistical detectability, while structurally marginalized groups remain outside the scope of formal interventions. For example, a hiring algorithm might reveal measurable gender disparities but fail to register racialized harms because of small sample sizes or coarse racial categories, leading gender to be treated as the "relevant" axis of disadvantage while racial inequities, though structurally embedded, go unaddressed. This logic implicitly ranks oppressions by their statistical measurability, undermining intersectionality's core rejection of the idea that any single axis of oppression is more fundamental than others. Similar dynamics have long shaped intersectionality debates, where attempts to capture multiple forms of discrimination produced terms like *double* discrimination or *double* disadvantage [111]. Yet these enumerative strategies often culminated in a vague "etc.," signaling both conceptual exhaustion and discomfort with the complexity of intersecting harms. Crenshaw [2] cautioned against this bureaucratic narrowing, arguing that treating intersectionality as a call to merely *expand* the list of protected classes collapses its critical analytic force into a classificatory exercise. Intersectionality's aim, by contrast, is to interrogate which relations of power are most *salient* in a given *context*, not to rank or enumerate identities.

A substantive approach to intersectional algorithmic fairness *rejects* the ranking, weighing, or hierarchical ordering of oppression and *resists* selecting categories based on data availability, institutional recognition, or algorithmic legibility. This requires shifting from statistically convenient attributes to those that are contextually and structurally *salient*. Rather than defaulting to legally or institutionally predefined attributes, a more substantive approach draws on socio-historical contexts to understand how categories function within power relations. This involves asking which identities matter in a given setting, how they have been historically constructed, and who decides their relevance. Klein und D'Ignazio [30] argue that a reversal of logic must guide such efforts, thus not beginning with the data at hand but with the questions that are missing. This means fairness assessments should start by identifying which constellations of disadvantage are most silenced or excluded, not which ones are most detectable, thus *centering the lived realities of the most marginalized* without implying that their oppression is more fundamental or should be treated as a new universal baseline. Crenshaw [2] notes the practical value of such an approach: „It is somewhat ironic that those concerned with alleviating the ills of racism and sexism should adopt such a top-down approach to discrimination. If their efforts instead began with addressing the needs and problems of those who are most disadvantaged and with restructuring and remaking the world where necessary, then others who are singularly disadvantaged would also benefit. In addition, it seems that placing those who are currently marginalized in the center is the most effective way to resist efforts to compartmentalize experiences and undermine potential collective action.“ Thus, centering the lived experiences of those affected is essential, regardless of whether discrimination can be neatly attributed to race, gender, or another singular axis.

Similar to the previous desideratum, a key step is enabling individuals to *self-identify* the social categories most relevant to their lived experience, instead of constraining them to institutionally predefined labels [112]. Such user-driven identity representation, combined with participatory processes in dataset construction and evaluation, can enhance both fairness and epistemic legitimacy by granting affected communities greater agency. Recent methodological advancements provide practical tools for this reorientation. For example, Decker et al. [104] propose the use of personas, contextually grounded, multidimensional profiles based on empirical research, as a way to capture complex social identities in algorithmic assessments. Similarly, Belitz et al. [113] advocate for a bottom-up, mixed-methods framework that identifies salient identity dimensions through qualitative interviews and tools like the Twenty Statements Test (TST). Their approach challenges unidimensional demographic categories and allows individuals to describe themselves in ways that reflect their lived experience. This, they argue, enhances not only the validity of fairness assessments but also the agency of affected communities. Public participation in dataset curation and fairness evaluation ensures that those subject to algorithmic decision-making have a voice in how their identities are represented and how fairness is defined [25]. *Technically*, context-aware data augmentation methods, including the careful use of generative AI, offer a promising means of addressing representational gaps, provided transparent normative commitments guide them and do not abstract away from the structural conditions they aim to model [114]. Additionally, adaptive fairness techniques such as dynamic weighting that respond to socio-political relevance can help ensure that interventions remain sensitive to varying contexts. Crucially, calls to "collect more data" from underrepresented groups must be *tempered by attention to the risks of surveillance, consent, and data exploitation* [85, 115]. A substantive intersectional approach requires not simply more data but different governance logics for deciding whether data should be collected at all. Recent work in critical data governance suggests two minimal safeguards: first, centering community consent, co-determination, and refusal as legitimate outcomes [87, 98, 116]; and second, adopting purpose limitation and harm-reduction frameworks that evaluate whether increased data granularity would meaningfully advance justice or instead amplify vulnerability [117, 118]. These principles do not resolve all trade-offs but offer a structured way of thinking about them: if the risks of extraction outweigh the fairness benefits, the normative commitment should shift from fixing data scarcity to redesigning the system so that it does not require vulnerable groups' data to function at all.

## 4.3 Overcoming The Lack of Attention to Socio-Technical Systems

While the previous section dealt with desiderata that pertain to unbreaking social categories, this section is committed to context, impact, and power surrounding algorithmic systems.

**Desideratum 6.** *Mapping power and domination structures.*

Intersectionality, unlike diversity or heterogeneity frameworks, places power relations at the *center* of analysis. As we have previously touched upon, categories such as race, class, gender, sexuality, disability, ethnicity, nationality, religion, and age are not merely *descriptive* but instead derive their meaning from historically situated relations of power and oppression, including racism, sexism, heterosexism, and class exploitation [4]. Algorithmic systems are *developed and deployed within* these structures, and their outputs can reproduce or intensify existing hierarchies that govern access to resources, recognition, and representation (see also Section 3.3.3). Collins' [40] four "domains of power" provide a valuable lens for examining these dynamics in algorithmic contexts [27]. Consider the case of a German-language proficiency model used in job placement, which may classify Black migrant women as having "low integration potential." In the *structural* domain, such systems encode institutional biases by presenting "language competence" as a neutral prerequisite for employability, while in practice reflecting racialized and gendered assumptions about citizenship and economic value. In the *disciplinary* domain, they may penalize speech patterns that deviate from standardized models, misinterpreting accent or code-switching as signs of deficiency. In the *hegemonic* domain, the very ideal of the "proficient speaker" reinforces cultural narratives of belonging tied to whiteness and nativeness. Finally, in the *interpersonal* domain, these classifications shape everyday interactions, as employers or case workers rely on algorithmic scores to judge integration potential, filtering out candidates whose intersecting identities diverge from dominant expectations. Crucially, the operation of these domains is *context-specific*. While in some cases "language competence" may constitute a legitimate skill to be hierarchically assessed, in others it functions as a proxy for exclusion.

A substantive approach to intersectional algorithmic fairness requires practices that interrogate how structural conditions shape algorithmic outcomes, without implying that technical actors must resolve all underlying social problems. Rather, their responsibility is twofold: to remediate harms for which algorithmic systems are a proximate cause, and to identify limits, refusing or redirecting deployment when harms stem primarily from broader institutional or societal structures [32]. *Context-specific power mapping conducted early in the problem-framing stage* can help in identifying the actors, institutions, and socio-political forces shaping algorithmic outcomes. This entails tracing upstream influences, such as data ownership, collection protocols, and institutional mandates, as well as examining downstream effects on access to resources, recognition, and representation. Thus, socio-technical audits should be institutionalized, integrating *quantitative subgroup disparity analysis with qualitative case studies*, ensuring that statistical findings are grounded in the realities of affected groups. *Community governance structures* can help in ongoing monitoring, redress mechanisms, and iterative design adjustments. Crucially, a substantive approach highlights that there is *no shared language of ethical concerns* across the human species [119]. Thus, considering the prevailing homogeneity of the machine learning field, a substantive approach would benefit from integrating perspectives that challenge dominant epistemologies and power paradigms. *Decolonial AI* scholarship demonstrates how colonial legacies shape notions of intelligence, value, and legitimacy in AI development, arguing for practices that confront these histories and foreground community agency [100, 120, 121]. More radical proposals, such as *data resurgence*, call for community-driven reinvestment in marginalized identities as a counter to technosolutionist narratives of progress [101]. Relational autonomy frameworks, such as Ubuntu, further illustrate how alternative value systems can reorient design: rather than centering individualistic assumptions about agency, Ubuntu emphasizes interdependence, mutual obligation, and collective flourishing [122]. Operationally, this implies co-determination of system goals with affected communities, communal rather than individual utility metrics, and shared accountability structures analogous to those developed in Indigenous data sovereignty governance [98, 116].

**Desideratum 7.** *Acknowledging that small actions can have a significant impact – and a distinct impact for distinct groups.*

Seemingly small actions within algorithmic systems, such as minor modifications to system parameters, model thresholds, or data inputs, can produce significant and distinct effects across different social groups (see also Section 3.3.4). This dynamic parallels the phenomenon of microaggressions [123], where ostensibly small or subtle acts can accumulate into profound harm over time. Acknowledging *social context* within intersectionality entails understanding that identical technological interventions may produce markedly divergent outcomes across diverse populations. For example, a recent evaluation of seat allocations in a labor market program found distribution to be "fair" by applied criteria. Yet, post-intervention unemployment rates remained stagnant because the model underestimated women's long-term unemployment risk, thus masking gender-specific structural vulnerabilities [124]. This phenomenon is particularly salient for multiply marginalized groups, who often experience compounded and nuanced forms of disadvantage. As Selbst et al. [31] caution, two design pitfalls are central here: the *framing trap* fails to situate algorithms within the broader social systems they affect. Additionally, the *ripple effect trap* overlooks how small technical changes can propagate through social networks and institutional processes, generating unintended and inequitable consequences.

A substantive approach, therefore, requires modelling and auditing not only immediate outputs but also longitudinal, context-specific impacts of algorithmic interventions. Green und Hu [125] highlight the necessity of maintaining a forward-looking perspective that captures evolving impacts, rather than limiting assessment to snapshots in time. Methodologically, Liu et al. [126] operationalize this through predictive modeling of variable changes induced by algorithmic deployment. Zezulka und Genin [124] model counterfactual outcomes to evaluate long-term impacts.

**Desideratum 8.** *Aligning purpose with context and impact of actions.*

Requiring that the intended purpose of an algorithmic system be rigorously aligned with the social context in which it is deployed, as well as with the actual downstream impacts of its outputs and interventions, means rejecting a "one-size-fits-all" deployment model in favor of context-sensitive design, fine-tuning, and, in some cases, deciding *not to deploy machine learning at all*. Practitioners risk harming underrepresented groups when they train models on data that did not fit the target population in the first place place [127, 128], apply algorithmic systems developed for a specific purpose on social environments they are not suited for [31, 129], or not consider approaches to social problems that do not involve technical solutions at all [31].

A substantive approach, therefore, begins with a problem-framing stage that interrogates the root causes of the issue at hand, evaluates whether an algorithmic system is the most appropriate intervention, and considers non-technical, community-driven alternatives [130]. Fairness objectives must be derived from, and remain accountable to, the overarching social goals of the intervention, ensuring that technical optimization does not drift from substantive justice aims [131]. Such an approach demands iterative testing in diverse, context-specific conditions, with explicit attention to how identical model behaviors may generate radically different consequences across populations. In practice, this entails integrating impact assessments that capture both quantitative performance metrics and qualitative accounts of lived experience, and updating or withdrawing systems when harms are identified. A substantive intersectional approach does not require algorithmic systems to resolve structural oppression; rather, it requires that their deployment not reinforce or intensify it. Accordingly, algorithmic interventions must include an explicit evaluation of whether the system is an appropriate instrument for the task at hand. Following prior work on algorithmic impact assessments and socio-technical audits, we consider non-deployment a necessary option when foreseeable harms stem primarily from entrenched structural conditions that the system cannot remediate, or when mitigation would require epistemic or material resources beyond the scope of algorithmic intervention. Non-deployment in this sense is not a rejection of technological innovation but a recognition of the limits of algorithmic governance and a commitment to preventing the reproduction of inequity where technical fixes are insufficient.

**Desideratum 9.** *Explicitly considering privileges and not only disadvantages.*

As discussed in desiderata 4 and 5, categorization in algorithmic systems is itself an exercise of power, often *privileging dominant perspectives* while marginalizing others [105]. Intersectionality, however, addresses not only the dynamics of disadvantage but also the often-overlooked operations of *privilege* (see also Section 3.3.2). Whereas desideratum 2 examines positionality on the side of system designers, this desideratum focuses on the privileges and structural advantages embedded within the identities of those *subjected* to algorithmic decision-making. Flagg [132] describes the pervasive "tendency of whites not to think about whiteness or about norms, behaviors, experiences, or perspectives that are white-specific." Collins' [40] intersectional matrix of domination underscores that every individual simultaneously receives both benefits and harms, penalties and privileges, depending on the social axes in question. Ignoring privilege, thus leaving the structural advantages of dominant groups such as whiteness or maleness unexamined, produces an incomplete understanding of social hierarchies and limits the capacity to address them. This unexamined normativity perpetuates what McIntosh [133] calls the invisible knapsack: an unearned set of assets, expectations, and immunities so deeply embedded in social systems that they remain invisible to those who benefit from them. When fairness assessments focus only on measuring harm, they risk reinforcing the idea that privileged groups are the natural baseline for comparison [134]. In addition to that, Crenshaw [2] observed that social categories frequently center on the most privileged members within marginalized groups[9] (e.g., white women within gender discrimination or the most privileged Black individuals within racial discrimination), thereby rendering invisible the norms that advantage these groups.

In algorithmic fairness, this neglect manifests as a prevailing narrative centered almost exclusively on harm to disadvantaged groups, while the structural advantages of dominant groups remain hidden. A substantive intersectional approach moves beyond deficit-focused interventions to a dual analysis of advantage and disadvantage, making advantages empirically visible and incorporating them into both diagnosis and remediation. This involves conducting structured audits to identify which social locations are most *advantaged* in the dataset, examining how institutional trust, access to resources, and historic inclusion in data collection may *disproportionately inflate performance metrics for privileged groups*. It also requires pairing harm-based metrics, such as false negative rates for disadvantaged groups, with privilege-based metrics that detect systematic overestimation of positive outcomes for dominant groups [e.g., 135]. Furthermore, counterfactual deployment scenarios can be used to simulate contexts where privileged group advantages are neutralized, for example, by removing features or proxies that disproportionately benefit the dominant group, and comparing these outputs with actual results to reveal the extent to which disparities are driven by privilege rather than relevant skill, need, or merit.

[9] She illustrates this in her ‚but for' notion: Addressing women often means people who are not marginalized at all, ‚but for' their gender.

### 4.4 Acknowledging Transformative Potential

This section discusses only one desideratum, albeit an important one.

**Desideratum 10.** *Recognizing algorithmic systems' potential for intersectionality beyond critique.*

While much of the discourse on intersectional algorithmic fairness rightly centers on critique, such as exposing algorithmic harms, systemic biases, and epistemic injustices, intersectionality also calls for proactive, emancipatory engagement. As Collins [33] emphasizes, intersectionality is not merely a diagnostic lens but a methodology for generating new knowledge that foregrounds the lived experiences of multiply marginalized communities. As Crenshaw [2] emphasizes, the challenge of intersectional discrimination often lies in its invisibility to dominant perspectives. Algorithmic systems, particularly when applied to large, complex datasets, can help identify previously unrecognized discriminatory effects. For instance, studies on spatialized discrimination, such as delivery services prioritizing ZIP codes, *reveal* how ostensibly neutral algorithms can disadvantage low-income, racialized communities; thus, algorithmic systems can not only uncover these inequities but also guide resource allocation and policy intervention accordingly.

To realize this potential, algorithmic systems can be deployed not solely as a diagnostic tool but as a vehicle for actively shaping equitable decision-making infrastructures. Klein und D'Ignazio's [30] notion of "data co-liberation" reframes the data pipeline from extraction to empowerment, demanding that algorithmic systems serve communities rather than exploit them. Zhang [136] offers a compelling model of such constructive application through the concept of *affirmative algorithms*, drawing on Elizabeth Anderson's theory of relational equality. She proposes interventions that actively support marginalized groups' substantive capabilities. In pretrial risk assessment, for example, such algorithms can adjust outputs to bolster Black defendants' access to justice affirmatively, recognizing that formal equality in an unequal society often perpetuates injustice. These approaches demonstrate that algorithmic systems can play a constructive role in the realization of intersectional justice. This requires prioritizing interventions that deliver tangible benefits to historically marginalized groups and reconfiguring algorithmic systems as a site for participatory, affirmative, and emancipatory practice. Not least the emergence of dedicated research spaces such as FAccT, AIES, and EWAF already reflects the field's recognition of these stakes, fostering interdisciplinary scholarship that resists narratives of algorithmic neutrality and centers feminist, decolonial, and anti-racist perspectives. Taken together, these insights position algorithmic systems not only as an object of critical reflection but as a tool for social transformation, capable of operationalizing the social justice aims at the heart of intersectional thought.

## 5 Conclusion

Drawing on feminist scholarship, we have applied intersectionality to algorithmic systems and derived ten desiderata that foster a more substantive engagement with it. We term this approach *Substantive Intersectional*

*Algorithmic Fairness*, enriching Green's [24] substantive algorithmic fairness with intersectional feminist perspectives to highlight their value for enhancing social justice, and contrasting it to the predominantly formal approaches prevalent in technical literature. Rather than providing a fixed operationalization, our desiderata are designed to offer actionable guidance: making assumptions of neutrality transparent, questioning the rigid use of protected attributes, situating algorithmic systems within broader social and power contexts, and recognizing their potential to promote social justice. Central to this endeavor is the inclusion of perspectives from multiply marginalized groups, whose experiences illuminate entrenched hierarchies and inform strategies for transformative change. At the same time, we recognize the limitations of our proposal. On the one hand, the desiderata are *minimal requirements* rather than comprehensive solutions, but on the other hand, their implementation is often complex and costly. Future work must therefore include feasibility studies and empirical cases to examine how these requirements can be realized in practice. In this regard, we highly value the contemporaneous contribution by Vethman et al. [29], who align their recommendations with practitioners' perspectives, and we view such plurality of approaches as vital for advancing the field. A key takeaway of our work is that complex social science concepts must not be appropriated by technical sciences in ways that reduce them to oversimplified endpoints. While engagement across domains is highly valuable, and some degree of simplification is unavoidable, it is crucial that concepts travel carefully between disciplines. Our desiderata seek to bridge these discourses and to remind us that algorithmic fairness cannot be pursued in isolation from its substantive social impact. Finally, technical constraints, such as the statistical exclusion of small groups, demand reflection on the very meaningfulness of algorithmic interventions in themselves. Substantive Intersectional Algorithmic Fairness calls not merely for technical fixes, but for principled, context-sensitive decisions about whether, when, and how algorithmic systems should be deployed. Our approach is grounded in intersectional feminist theory for its capacity to connect lived experience with structural analysis and its explicit commitment to redressing imbalances of power. At the same time, we acknowledge that intersectionality is not the only framework for analyzing structural inequality, and we welcome parallel and alternative approaches as part of a plural and critical research agenda. As McFadden und Alvarez [137] remind us, understanding is a necessary prerequisite for action. With this contribution, we aim to foster such understanding and invite critical engagement to refine and expand these recommendations further.

# Statements and Declarations

## Funding

This work was funded by the Federal Ministry of Education and Research (BMBF) and the Ministry of Culture and Science of the German State of North Rhine-Westphalia (MKW) under the Excellence Strategy of the Federal Government and the Länder.

## Competing Interests

The authors have no competing interests to declare that are relevant to the content of this article.

## Ethical approval

This article does not report any studies with human participants or animals performed by the authors.

## Informed consent

Not applicable.

## Data availability

No new datasets were generated or analyzed in this study.